\documentclass{article}

\usepackage[utf8]{inputenc}
\usepackage[T1]{fontenc}

\usepackage{amsmath}
\usepackage{amssymb} 
\usepackage{xfp}
\usepackage{amsthm}
\usepackage{bm} 

\newtheorem{theorem}{Theorem}
\newtheorem{remark}{Remark}

\usepackage{natbib} 

\usepackage{tikz}
\usetikzlibrary{decorations.pathmorphing,math}
\usepackage{circuitikz}

\usepackage{hyperref}

\hypersetup{
  colorlinks = true,
  linkcolor = blue,
  citecolor = blue,
  urlcolor = magenta,
  pdftitle = {Guaranteed Time Control using Linear Matrix Inequalities},
  pdfauthor = {Campos, Quadros, Frezzatto, Mozelli, Nguyen},
}

\renewcommand{\vec}[1]{\bm{#1}}

\makeatletter
\newcommand{\pushright}[1]{\ifmeasuring@#1\else\omit\hfill$\displaystyle{#1}$\fi\ignorespaces}
\makeatother

\begin{document}

\title{Guaranteed Time Control using Linear Matrix Inequalities}
\author{Víctor Costa da Silva Campos \\ Department of Electronics Engineering\\ Universidade Federal de Minas Gerais
 \and Mariella Maia Quadros \\  Area of Control and Industrial Processes \\ Instituto Federal de Minas Gerais
 \and Luciano Frezzato \\ Polytechnic School\\ University of São Paulo
 \and Leonardo Amaral Mozelli \\ Department of Electronics Engineering\\ Universidade Federal de Minas Gerais
 \and Anh-Tu Nguyen \\ Laboratory LAMIH UMR CNRS 8201, INSA  Hauts-de-France\\ Université Polytechnique Hauts-de-France}

\maketitle

\begin{abstract}
This paper presents a synthesis approach aiming to guarantee a minimum upper-bound for the time taken to reach a target set of non-zero measure that encompasses the origin, while taking into account uncertainties and input and state constraints.
This approach is based on a harmonic transformation of the Lyapunov function and a novel piecewise quadratic representation of this transformed Lyapunov function over a simplicial partition of the state space.
The problem is solved in a policy iteration fashion, whereas the evaluation and improvement steps are formulated as linear matrix inequalities employing the structural relaxation approach.
Though initially formulated for uncertain polytopic systems, extensions to piecewise and nonlinear systems are discussed.
Three examples illustrate the effectiveness of the proposed approach in different scenarios.
\end{abstract}

\section{Introduction}

    Time constraints are relevant in both controller and observer performances. For control, the settling time is an important specification that enables faster transient responses and more robustness to disturbances, which can be motivated for safety or productivity reasons. For observation, faster convergence can be critical to using the estimated states in real-time applications. Therefore, time constraints are present in many engineering problems as shown in \cite{LiuJAS2022} and references therein.
    
    When dealing with precisely known asymptotically stable linear systems, with no regards to state constraints, the time-optimal control problem is known to be solved by employing the \emph{bang-bang principle} \citep{LaSalle1960}. In more general settings, the \emph{Kruzkov Transformation} can be employed to transform the \emph{Hamilton-Jacobi-Bellman} (HJB) equations associated with the time-optimal control problem into similar HJB equations as discounted optimal control problems, which can be solved numerically \citep{Bardi1997,Falcone2013}. However, this approach can be trickier when the system under consideration does not satisfy certain regularity conditions, usually \emph{small-time controllability} of the target set and Soner's inward pointing condition, since the viscosity solution of the HJB can be discontinuous \citep{Altarovici2013}. 
    
    Instead of directly solving the time-optimal control problem, finite-time stability \citep{Bhat2000} defines a set of Lyapunov conditions that, if satisfied, ensures the origin of the state space is reached in a finite amount of time (in contrast to asymptotic stability which only requires the origin to be reached as $t \rightarrow \infty$). Even though these are useful analysis conditions, they are often hard to employ in a constructive manner for control laws. As such, it is common to employ Implicit Lyapunov Function (ILF) designs \citep{Polyakov2015}, or Prescribed Performance Control (PPC) \citep{Guo2022} in order to impose finite-time stability on the closed loop, with the main disadvantage of these approaches being the fact that they require a specific structure upon the system's dynamics.
    
    Many approaches disregard the fact that most applications have constraints upon the system's state and input, as well as uncertainties on the available model. Throughout the last 30 years, robust control conditions based on Linear Matrix Inequalities (LMIs) \citep{Boyd1994} have been shown to be particularly useful in this regard, specially if the uncertainties can be written as a convex set (leading to a polytopic representation). An approach that is already well-developed and employed in the literature is to make use of a saturation function in the control law, and ensuring the closed loop's stability inside of a predetermined level set of a Lyapunov function \citep{Tarbouriech2011}. Even though the existence of a quadratic Lyapunov function is necessary and sufficient for the stability of precisely known linear time-invariant systems, it is a well-known fact that it is a source of conservativeness when dealing with uncertain \citep{Ramos2022} or nonlinear systems \citep{Nguyen2019}.
    
    An interesting approach in that regard is the use of Piecewise Quadratic Lyapunov functions \citep{Johansson1998}, which, given enough partitions of the state space, are capable of approximating any nonlinear function. As shown by \cite{Rantzer2000}, they can be employed to approximate upper-bounds of optimal control problems, though this approach is only valid for the largest Lyapunov level set contained in the analysis region (which can be hard to find in this scenario). Building on this idea, \cite{Jiang2015} propose a relaxed optimal control problem (for polynomial systems), and a \emph{Policy Iteration} solution based on \emph{sum of squares} (SOS), which is capable of finding a global suboptimal solution for the relaxed optimal control problem given an initial policy/control law. Its main disadvantage lies on the fact that it does not take into account state and input constraints.
    
    Inspired by these works, this paper proposes a Guaranteed Time Control approach, in which the goal is to minimize an upper-bound of the time taken to reach a desired region that covers the origin of the state space, while respecting bounds on the state space and control inputs and ensuring asymptotic stability of the origin. In order to do so, we formulate a theorem with Lyapunov conditions based on the \emph{Harmonic Transformation} \citep{Campos2025}, which, if satisfied, not only provides an upper-bound on the time taken to reach the desired region, but also an estimate of the origin's domain of attraction. In order to make use of these new conditions, we propose a novel representation for piecewise quadratic Lyapunov functions, based on a simplicial partition of the state space, and a \emph{Policy Iteration} approach based on Linear Matrix Inequalities relying on the structural relaxation approach \citep{kim2024,Campos2025b} to introduce local information into the conditions. A remark details how the proposed conditions can be easily modified to deal with piecewise polytopic systems, and nonlinear systems represented by Takagi-Sugeno fuzzy models \citep{Takagi1985}. Three examples are presented to illustrate the proposed approach and compare it against the literature.
    
    \subsection*{Notation}
    We represent matrix, vectors and scalar variables by uppercase, \textbf{bold text lowercase}, and lowercase letters respectively. Though an abuse of notation, some scalar functions will be represented by uppercase variables, either to avoid confusion, or due to that being the standard use in the literature. Given a set $\mathcal{A}$, its boundary is represented by $\partial\mathcal{A}$. Given matrix $M \in \mathbb{R}^{r \times p}$, $\vec{m}_j$ represents its $j$-th column, and $m_{ij}$ represents the element in row $i$ and column $j$. For matrices, $S \succeq 0$ ($\preceq 0$) describes that the elements of the matrix are non-negative (non-positive). For symmetric matrices, $M > 0$ ($< 0$) describes that matrix $M$ is positive (negative) definite, and $M \geq 0$ ($\leq 0$) is positive (negative) semidefinite. $M^T$ represents the transpose of a matrix, $M^{-1}$ its inverse, and $M^{-T} = \left(M^T\right)^{-1}=\left(M^{-1}\right)^T$ the inverse of the transpose. When describing a symmetric matrix, the symbol $\ast$ represents block transpose terms that can be inferred from the symmetry. Throughout this paper, different convex sums will be employed in different contexts (with different convex weights), and the short notation $S_\alpha = \sum_{i = 1}^r \alpha_i S_i$ is employed. When dealing with a set of indices up to $s$ we will make use of the representation $\mathcal{I}_s = \{i \in \mathbb{N} |\ 1 \leq i \leq s\}$. Time dependency will usually be dropped, and will only be shown explicitly when deemed necessary.

\section{Methodology}
    Consider an uncertain polytopic system described by
    \begin{align}
        \dot{\vec{x}} = \sum_{i = 1}^r \alpha_i \left(A_i \vec{x} + B_i \vec{u} \right) = A_\alpha \vec{x} + B_\alpha \vec{u}, \label{eq:sys}
    \end{align}
    with $\vec{x} \in \mathcal{X} \subset \mathbb{R}^n$ and $\vec{u} \in \mathcal{U} \subset \mathbb{R}^m$ representing the states and the control inputs from bounded sets $\mathcal{X}$ and $\mathcal{U} = \{\vec{u} \in \mathbb{R}^m |\; \underline{u}_i \leq u_i \leq \bar{u}_i, \forall i = \mathcal{I}_m\}$, respectively; $r$ the number of vertices in the polytopic representation, $A_i \in \mathbb{R}^{n \times n}, B_i \in \mathbb{R}^{n \times m}$ the matrices representing the system's dynamics, and $\bm{\alpha}^T = \begin{bmatrix} \alpha_1 & \dots & \alpha_r \end{bmatrix}$ the uncertain polytopic convex weights satisfying $\alpha_i \geq 0, \forall i$ and $\sum_{i=1}^r \alpha_i = 1$. Our goal is to find a control law capable of driving the largest possible subset of $\mathcal{X}$ towards a set $\mathcal{X}_g$, that contains the origin, in the shortest time and without violating the bounds imposed by  $\mathcal{U}$. To that end, consider the following theorem.

    \begin{theorem} \label{thm:harmonic_lyap}
        Consider that there exists a positive definite function $\bar{V}(\vec{x}) : \mathcal{X} \rightarrow \mathbb{R}$ with the boundary conditions
        \begin{equation}
            \left\{\begin{array}{ll} \bar{V}(\vec{x}) = 0, & \vec{x} = 0, \\ \bar{V}(\vec{x}) \geq 1, & \vec{x} \in \partial \mathcal{X},\end{array}\right.
        \end{equation}
        
        whose time derivative satisfies
        \begin{equation}
            \left\{\begin{array}{ll}\dot{\bar{V}}(\vec{x}) + \left(1 - \bar{V}(\vec{x})\right)^2 \leq 0, &\textrm{ if } \vec{x} \not\in \mathcal{X}_g, \\
            \dot{\bar{V}}(\vec{x}) < 0, & \textrm{ if } \vec{x} \in \mathcal{X}_g.
            \end{array}\right.
        \end{equation}
        Then, the origin is asymptotically stable, the \emph{strict} 1-sublevel set of $\bar{V}(\vec{x})$, $\Omega_{\bar{V}_1} = \{\vec{x} \in \mathcal{X} |\ \bar{V}(\vec{x}) < 1\}$, is an estimate of the origin's domain of attraction, and any point $\vec{x} \in \Omega_{\bar{V}_1}$ is guaranteed to reach the set $\mathcal{X}_g$ in finite time, with an upper-bound estimate of the time taken to reach the set given by $\dfrac{\bar{V}(\vec{x})}{1 - \bar{V}(\vec{x})}$.
    \end{theorem}

    \begin{proof}
        The conditions of the theorem ensure that $\bar{V}(\vec{x})$ is positive definite, with a negative definite time-derivative for $\vec{x} \in \Omega_{\bar{V}_1}$, which ensures that the origin of the state is asymptotically stable. Since the theorem requires that $\bar{V}(\vec{x}) \geq 1$ on the boundaries of the $\mathcal{X}$ analysis region, the $\Omega_{\bar{V}_1}$ set is entirely contained within the region, which, together with the fact that $\bar{V}(\vec{x})$ is a Lyapunov function, ensures that $\Omega_{\bar{V}_1}$ is positively invariant and can be used as an estimate of the origin's domain of attraction. For any state in $\Omega_{\bar{V}_1} \setminus \mathcal{X}_g$, $\bar{V}(\vec{x}) < 1$ and it follows that
        \begin{align}
            \dfrac{\dot{\bar{V}}(\vec{x})}{\left(1 - \bar{V}(\vec{x})\right)^2} + 1 \leq 0.
        \end{align}
        If we define $V(\vec{x}) = \dfrac{\bar{V}(\vec{x})}{1 - \bar{V}(\vec{x})}$, we have that $\dot{V}(\vec{x}) = \dfrac{\dot{\bar{V}}(\vec{x})}{\left(1 - \bar{V}(\vec{x})\right)^2}$ and 
        \begin{align}
            &\dot{V}(\vec{x}) + 1 \leq 0, \\
            &\dot{V}(\vec{x}) \leq -1.
        \end{align}
        From the Comparison Lemma \cite[Lemma 3.4]{Khalil2002}, it follows that, if we integrate both sides from time $t$ to $t + \Delta t$, with $\Delta t$ the time taken to reach the set $\mathcal{X}_g$ from $\vec{x}$,
        \begin{align}
            &V(\vec{x}(t + \Delta t)) \leq V(\vec{x}(t)) - \Delta t 
        \end{align}
        with $V(\vec{x}(t + \Delta t))$ the value that function $V(\vec{x})$ takes at the point the state is entering this region. By noting that the transformation from $\bar{V}(\vec{x})$ to $V(\vec{x})$ maps $[0,1)$ to $[0,\infty)$, and in turn guarantees that $V(\vec{x}(t + \Delta t)) \geq 0$, it follows that
        \begin{align}
            \Delta t \leq V(\vec{x}) = \dfrac{\bar{V}(\vec{x})}{1 - \bar{V}(\vec{x})},
        \end{align}
        which concludes the proof.
    \end{proof}

    \begin{remark}
        In Theorem \ref{thm:harmonic_lyap}, $\bar{V}(\vec{x})$ is the \emph{Harmonic Transformation} \citep{Campos2025} of the function $V(\vec{x})$, which is a function upper-bounding the time taken to reach region $\mathcal{X}_g$. The \emph{Kruzkov Transformation} \citep{Bardi1997,Falcone2013} is usually employed in time-optimal control with the same goal of mapping $[0,\infty)$ to $[0,1)$ (allowing the representation of possible infinite values by using 1). However, as demonstrated in \citep{Campos2025}, even though the \emph{Kruzkov Transformation} usually leads to simpler conditions, the \emph{Harmonic Transformation} behaves better numerically, motivating its choice for the current work.
    \end{remark}

    Theorem \ref{thm:harmonic_lyap} is an analysis tool capable of estimating both a domain of attraction as well as the time to reach the set $\mathcal{X}_g$, and any function satisfying its conditions would give us a time upper-bound. In trying to find the smallest upper-bound, we can use the conditions of the theorem as constraints and the integral of the $\bar{V}(\vec{x})$ function over the state space ($\int_\mathcal{X} \bar{V}(\vec{x}) \vec{dx}$) as the cost in an optimization problem. Similar to other approaches in the literature \citep{Jiang2015,Saenz2021,Pakkhesal2024}, employing optimization algorithms to solve for the conditions in Theorem \ref{thm:harmonic_lyap} can be seen as a sort of Approximate Dynamic Programming (ADP), since the upper-bound solutions that satisfy the conditions in the theorem can be seen as supersolutions to the corresponding \emph{harmonically transformed} Hamilton-Jacobi-Bellman equations for minimum-time problems \citep{Campos2025}, and solving for the minimum upper-bound can then be seen as trying to solve the \emph{Relaxed Optimal Control Problem} \citep[Problem 3.1]{Jiang2015}. Many of the optimization-based ADP approaches in the literature focus their attention on the \emph{global} solution, considering an unrestricted control set and quadratic costs in the control and states (leading to a closed form solution for the control law/policy given a certain Lyapunov/value function). They usually consider a polynomial representation, such that the optimization problem can be cast as a \emph{sum of squares} (SOS) optimization problem \citep{Jiang2015,Saenz2021,Pakkhesal2024}.

    \subsection{Piecewise Representation} \label{sect:piece}
    Unlike previous approaches in the literature, in this work we focus our attention to a \emph{local} constrained (both in states and control) setting. To accomplish this, we consider a piecewise quadratic representation for $\bar{V}(\vec{x})$ \citep{Johansson1998,Rantzer2000}, which is more directly amenable to deriving LMI conditions, and recasting our optimization as a Semi-Definite Programming Problem. However, similarly to \cite{Cabral2024}, we do not make use of the parametrization defined by the constraint matrices used in \cite{Johansson1998}, and present a \emph{new} parametrization for piecewise quadratic functions.

    \begin{figure}
        \centering
        \begin{tikzpicture}[]
            \tikzmath{\np = 7; \n1 = \np-1; \n2 = \np-2;}
            \fill[blue!10] (\fpeval{\n1/2},\fpeval{\n1/2}) ++(0,1) -- ++(-1,0) -- ++(0,-1) -- ++(1,-1) -- ++(1,0) -- ++(0,1) -- ++(-1,1);
            \foreach \i in {0,...,\n1} 
                \foreach \j in {0,...,\n2}
                {
                    \draw[red] (\i,\j) -- (\i,\j+1);
                }
            \foreach \i in {0,...,\n2} 
                \foreach \j in {0,...,\n1}
                {
                    \draw[red] (\i,\j) -- (\i+1,\j);
                }
            \foreach \i in {0,...,\n2} 
                \foreach \j in {0,...,\n2}
                {
                    \draw[red] (\i+1,\j) -- (\i,\j+1);
                }
            \foreach \i in {0,...,\n1} 
                \foreach \j in {0,...,\n1}
                {
                    \fill (\i,\j) circle (0.1cm);
                }
            \node at (\fpeval{\n1/2},\fpeval{\n1/2}) [above=4, right = 0.2] {\scriptsize $(0,0)$};
            \node at (\fpeval{\n1/2}-1,\fpeval{\n1/2}+1) [below=10, right = 10] {\small $\mathcal{X}_g$};
        \end{tikzpicture}
        \caption{Illustration of a Simplicial partition for $\mathcal{X}$ in the 2-dimensional case. As illustrated in the figure, we always consider that the origin, $(0,0)$ in this case, is a part of the grid of points used to partition the space. As illustrated by the blue colored region in the figure, we consider that every simplex which has the origin as a vertex is a part of the set $\mathcal{X}_g$ towards which we guarantee the upper-bound on convergence time.}
        \label{fig:triang_grid}
    \end{figure}
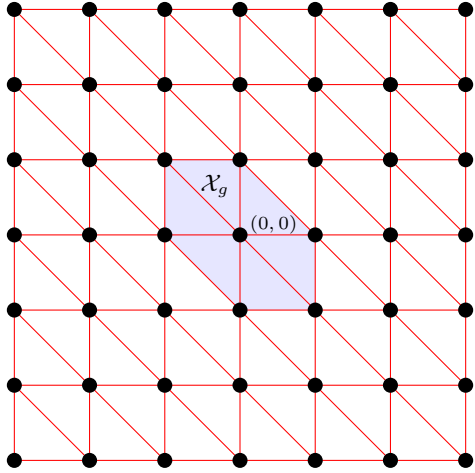

    Consider that we have a simplicial partition of the set $\mathcal{X}$,
    \begin{equation}
        \mathcal{X} = \bigcup_{q = 1}^{r_s} \vec{\Delta}_q, \label{eq:union_simplex}
    \end{equation}
    representing our local analysis region in the state space with $\vec{\Delta}_q$ representing each simplex in our partition, and $r_s$ the total number of simplices. This partition can always be found from a (possibly nonregular) grid of points $\Gamma = \{\vec{x}_1, \dots , \vec{x}_{r_p}\}$ in the state space, with $r_p$ the number of grid points, by using a \emph{Delaunay Triangulation} \citep{Delaunay1934}. We consider that the origin is always part of this generating grid, and that the set $\mathcal{X}_g$ is composed by the simplices which contain the origin as a vertex. This description is illustrated in Figure \ref{fig:triang_grid} for a 2-dimensional state space. In addition to this, the triangulation of the state space defines a mapping $\eta(q,j):\mathcal{I}_{r_s} \times \mathcal{I}_{n+1} \rightarrow \mathcal{I}_{r_p}$, which given indices $q$ and $i$ representing a simplex $\vec{\Delta}_q$ and one of the vertices of said simplex respectively, relates this vertex to a point on the grid $\Gamma$. Given this mapping, each simplex is defined as the convex hull of a set of $n+1$ vertex points, such that
    \begin{align}
        \vec{\Delta}_q &= \operatorname{co}\left(\vec{x}_{\eta(q,1)}, \dots, \vec{x}_{\eta(q,n+1)}\right), 
    \end{align}
    which implies that there exists $\vec{\beta} \in \mathbb{R}^{n+1}$ \emph{unique} convex weights that can be used to represent any point in the simplex from its vertices as $\vec{x} = \sum_{j = 1}^{n+1} \beta_j \vec{x}_{\eta(q,j)}$, with $\vec{\beta}^T = \begin{bmatrix} \beta_1 & \dots & \beta_{n+1}\end{bmatrix}$, $\beta_{j} \geq 0 \;\forall j, \; \sum_{j=1}^{n+1} \beta_j = 1$.

    For each point $\vec{x}_i$ in our grid, we define an affine function
    \begin{equation}
        \bar{v}_i(\vec{x}) = \vec{p}_{i}^T \begin{bmatrix} \vec{x} \\ 1 \end{bmatrix} = \vec{p}_i^T \bar{\vec{x}},
    \end{equation}
    with $\bar{\vec{x}}$ representing an augmented state vector, $\vec{p}_i \in \mathbb{R}^{n+1}$ the parameters defining the function, and $\vec{p}_{i(n+1)} = 0$ if $\vec{x}_i = \vec{0}$, such that $\bar{v}_i(\vec{0}) = 0$. Our Piecewise Quadratic function is given by the finite-element linear interpolation of these points, using our simplices as the finite-elements. Note that, for any point belonging to simplex $\vec{\Delta}_q$, we can write that
    \begin{align}
        \begin{bmatrix} \vec{x} \\ 1 \end{bmatrix} &= \begin{bmatrix} \vec{x}_{\eta(q,1)} & \dots & \vec{x}_{\eta(q,n+1)} \\ 1 & \dots & 1 \end{bmatrix} \begin{bmatrix} \beta_1 \\ \vdots \\ \beta_{n+1}\end{bmatrix},\\
        \bar{\vec{x}} &= \bar{X}_q \vec{\beta},
    \end{align}
    which allow us to calculate the convex weights by
    \begin{align}
        \vec{\beta} &= \bar{X}_q^{-1} \bar{\vec{x}}. \label{eq:beta_definition}
    \end{align}
    Note that, so long as there are no degenerate simplices in the triangulation (simplices whose hypervolume is zero), the inverse is always guaranteed to exist (since the hypervolume of simplex $\vec{\Delta}_q$ is proportional to $\det(\bar{X}_q)$). Inside of simplex $\vec{\Delta}_q$, the finite-element linear interpolation gives us
    \begin{align}
        V_q(\vec{x}) &= \sum_{j=1}^{n+1} \beta_j \vec{p}_j^T \bar{\vec{x}} = \begin{bmatrix} \beta_1 & \dots & \beta_{n+1} \end{bmatrix} \begin{bmatrix} \vec{p}_1^T \\ \vdots \\ \vec{p}_{n+1}^T \end{bmatrix} \bar{\vec{x}}, \\
        &= \vec{\beta}^T P_q^T \bar{\vec{x}} = \bar{\vec{x}}^T \bar{X}_q^{-T} P_q \bar{\vec{x}}, \\&= \bar{\vec{x}}^T \left(\dfrac{1}{2}\left(P_q^T \bar{X}_q^{-1} + \bar{X}_q^{-T} P_q\right)\right)\bar{\vec{x}}, \\
        V_q(\vec{x}) &=  \bar{\vec{x}}^T S_q \bar{\vec{x}},
    \end{align}
    with $P_q \in \mathbb{R}^{(n+1) \times (n+1)}$ a matrix whose columns correspond to the $\vec{p}_i$ vectors of each vertex composing simplex $\vec{\Delta}_q$, and 
    \begin{equation}
        S_q = \dfrac{1}{2}\left(P_q^T \bar{X}_q^{-1} + \bar{X}_q^{-T} P_q\right) \label{eq:Sq}
    \end{equation}
    the local quadratic parametrization of the Piecewise Quadratic function. With this description of the finite-element linear interpolation as a Piecewise Quadratic function, we can write that
    \begin{equation}
        \bar{V}(\vec{x}) = V_q(\vec{x}), \;\textrm{if } \vec{x} \in \vec{\Delta}_q. \label{eq:switch_V}
    \end{equation}
    It is important to note that, because this parametrization arises from a linear interpolation over the simplices, it is naturally continuous over the state space. The barycentric coordinates, which serve as the convex weights within each simplex, are uniquely defined as an affine function of the states (or a linear function over the augmented state vector $\bar{\vec{x}}$, as shown in \eqref{eq:beta_definition}). Moreover, on the shared faces of adjacent simplices, these coordinates reduce consistently to the barycentric coordinates associated with the vertices of the face, ensuring that the interpolation is continuous and uniquely defined across neighboring simplices.

    In this work, we consider that a control law, satisfying $\vec{u}(\vec{x}) \in \mathcal{U}, \forall \vec{x} \in \mathcal{X}$, is given by a (possibly discontinuous) Piecewise Linear function of the form
    \begin{equation}
        \vec{u}(\vec{x}) = \begin{bmatrix} K_q & \vec{k}_q \end{bmatrix} \begin{bmatrix} \vec{x} \\  1 \end{bmatrix} = \bar{K}_q \bar{\vec{x}}, \;\textrm{if } \vec{x} \in \vec{\Delta}_q,\label{eq:K}
    \end{equation}
    with $\bar{K}_q \in \mathbb{R}^{m \times (n+1)}$ attributing a gain for each simplex in the triangulation and $\vec{k}_{q} = 0$ if $\vec{\Delta}_q \in \mathcal{X}_g$, to ensure that $\vec{u}(\vec{0}) = \vec{0}$. With this control law, we can write that the closed loop dynamics are given by
    \begin{align}
        \dot{\bar{\vec{x}}} = \sum_{k=1}^r \alpha_k \bar{A}_{kq} \bar{\vec{x}},
    \end{align}
    with
    \begin{align}
        \bar{A}_{kq} = \begin{bmatrix} A_k + B_k K_q & B_k \vec{k}_q \\ \vec{0} & \vec{0} \end{bmatrix}. \label{eq:x_Akq}
    \end{align}

    Since we do not impose continuity upon the control law, the closed loop vector field may be discontinuous on the boundaries of the simplices. Ergo, the generalized solution of the closed loop dynamics will lie in the convex hull of both possible switching dynamics, and additional constraints need to be added to the optimization problem to ensure they will also satisfy our conditions. Furthermore, in order to impose that $\bar{V}(\vec{x}) \geq 1$ on the boundaries of $\mathcal{X}$, we also require a way to describe these boundary regions. Note that, in both cases (either boundaries with neighboring simplices or with the boundary of the analysis region), these regions can be described by the $n$ vertices composing the face, which can be seen as an $(n-1)$-dimensional simplex, $\vec{\delta}_\ell$. Considering that there are $r_f$ faces, representing all of the faces on the simplices of our triangulation, similarly to the simplex case, we define a mapping $\eta_f(\ell,j): \mathcal{I}_{r_f} \times \mathcal{I}_{n} \rightarrow \mathcal{I}_{r_p}$, which given an index $\ell$ representing a face $\vec{\delta}_\ell$, and an index $j$ representing one of its vertices, returns an index corresponding to a point on the grid $\Gamma$. Given this mapping, each face can be defined as the convex hull of a set of $n$ vertex points, such that
    \begin{align}
        \vec{\delta}_\ell &= \operatorname{co}\left(\vec{x}_{\eta_f(\ell,1)}, \dots, \vec{x}_{\eta_f(\ell,n)}\right),
    \end{align}
    which implies that there exists $\bar{\vec{\beta}} \in \mathbb{R}^{n}$ \emph{unique} convex weights that can be used to represent any point in the face from its vertices as $\vec{x} = \sum_{j = 1}^{n} \bar{\beta}_j \vec{x}_{\eta_f(\ell,j)}$, with $\bar{\vec{\beta}}^T = \begin{bmatrix} \bar{\beta}_1 & \dots & \bar{\beta}_{n}\end{bmatrix}$, $\bar{\beta}_{j} \geq 0 \;\forall j, \; \sum_{j=1}^{n} \bar{\beta}_j = 1$. We also consider that there exists a function $\nu(\ell) : \mathcal{I}_{r_f} \rightarrow \mathcal{I}_{r_s} \times \mathcal{I}_{r_s}$, which, given an index representing a face, returns the indices of the two neighboring simplices that share said face, or the same simplex twice (if the face is on the border of $\mathcal{X}$, it is only contained by a single simplex, which gets returned twice).

    \subsection{Policy Iteration}
    Like many other LMI based approaches, if we simply tried to find a set of solutions from Theorem \ref{thm:harmonic_lyap} that directly tried to solve for both the Lyapunov function $\bar{V}(\vec{x})$ and the control law, we would end up with a set of \emph{Bilinear Matrix Inequalities} (BMIs). In order to avoid that, as in many other works in the literature, we tackle the problem using \emph{Policy Iteration}, which amounts to the \emph{Policy Evaluation} and \emph{Policy Improvement} steps being repeated iteratively. \emph{Policy Evaluation} is a step in which, given a control law/policy, $\vec{u}(\vec{x})$ in \eqref{eq:K}, for the closed loop system, we try to find the smallest upper-bound Lyapunov function, $\bar{V}(\vec{x})$ in \eqref{eq:switch_V}. In contrast to this, \emph{Policy Improvement} is a step in which, given a Lyapunov function, $\bar{V}(\vec{x})$ in \eqref{eq:switch_V}, finds the control law that tries to maximize its decay. Whereas most optimization-based ADP approaches have a closed-form solution for the \emph{Policy Improvement} step \citep{Rantzer2000,Jiang2015,Saenz2021,Pakkhesal2024}, due to the fact that the control set $\mathcal{U}$ is bounded, and that our control law is parametrized by \eqref{eq:K}, we instead also propose a set of LMI conditions to solve this step. The conditions for both steps are presented in Theorems \ref{thm:policy_evaluation} and \ref{thm:policy_improvement} in the following.

    \begin{theorem}[Policy Evaluation] \label{thm:policy_evaluation}
    Consider the uncertain polytopic system given by \eqref{eq:sys}, with a control law given by \eqref{eq:K}. Consider that there are also matrix functions $\Lambda_\beta \in \mathbb{R}^{\frac{n(n+1)}{2} \times (n+1)}$ and $T_{\bar{\beta}} \in \mathbb{R}^{\frac{n(n-1)}{2} \times n}$ such that $\Lambda_\beta \vec{\beta} = 0, T_{\bar{\beta}} \bar{\vec{\beta}} = 0$. Given a scalar $\gamma > 0$, if there exist vectors $\vec{p}_\ell \in \mathbb{R}^{n+1}$ with $\vec{p}_{\ell (n+1)} = 0$ if $\vec{x}_\ell = \vec{0}$, non-negative symmetric matrices $W^1_{q}, W^2_{kq} \in \mathbb{R}^{(n+1) \times (n+1)}, \bar{W}_{kq} \in \mathbb{R}^{2(n+1) \times 2(n+1)}, E^1_q, E^2_{skq} \in \mathbb{R}^{n \times n}, \bar{E}_{skq} \in \mathbb{R}^{2n \times 2n}$, and matrices $N_{kq}, M_{q} \in \mathbb{R}^{(n+1) \times \frac{n(n+1)}{2}}, \bar{N}_{kq} \in \mathbb{R}^{2(n+1) \times n(n+1)}, G_q, H_{skq} \in \mathbb{R}^{n \times \frac{n(n-1)}{2}}, \bar{H}_{skq} \in \mathbb{R}^{2n \times n(n-1)}$ that minimize
    {\begin{equation}
        \dfrac{2}{(n+1)(n+2)n!}\left(\sum_{q = 1}^{r_s}\left|\det(\bar{X}_q)\right|\left(\sum_{i=1}^{n+1} \sum_{j=i}^{n+1} \bar{\vec{x}}_{\eta(q,i)}^T S_q \bar{\vec{x}}_{\eta(q,j)}\right)\right) \label{eq:int_numerical}
    \end{equation}}
    subject to
    \begin{align}
        &\bar{X}_q^T\left(S_q - \gamma \begin{bmatrix} I_n & \vec{0} \\\vec{0} & \vec{0} \end{bmatrix} \right)\bar{X}_q + M_{q} \Lambda_i + \Lambda_i^T M_{q}^T - W^ 1_{q} \geq 0, \\
        &\bar{X}_q^T\left(\bar{A}_{kq}^T S_q + S_q \bar{A}_{kq}\right)\bar{X}_q + N_{kq} \Lambda_i + \Lambda_i^T N_{kq}^T + W^ 2_{kq} < 0
    \end{align}
    for every $\Delta_q \in \mathcal{X}_g$, and
    \begin{align}
        &\bar{X}_q^T S_q \bar{X}_q + M_{q} \Lambda_i + \Lambda_i^T M_{q}^T - W^ 1_{q} \geq 0, \\
        &\dfrac{1}{2}\left(\bar{Q}^{(kq)} + \left.\bar{Q}^{(kq)}\right.^T \right) + \bar{N}_{kq} \bar{\Lambda}_i + \bar{\Lambda}_i^T \bar{N}_{kq}^T + \bar{W}_{kq} \leq 0
    \end{align}
    for every $\Delta_q \in \mathcal{X}\setminus\mathcal{X}_g$, with $i \in \mathcal{I}_{n+1}$, $k \in \mathcal{I}_{r}$, $S_q$ given by \eqref{eq:Sq},
        \begin{equation}
            \bar{X}_q = \begin{bmatrix} \vec{x}_{\eta(q,1)} & \dots & \vec{x}_{\eta(q,n+1)} \\ 1 & \dots & 1 \end{bmatrix},
        \end{equation}
        \begin{equation}
            \bar{Q}^{(kq)} = \begin{bmatrix}
                Q_{11}^{(kq)} & \dots & Q_{1(n+1)}^{(kq)} \\
                \vdots & \ddots & \vdots \\
                Q_{(n+1)1}^{(kq)} & \dots & Q_{(n+1)(n+1)}^{(kq)}
            \end{bmatrix}, 
        \end{equation}
        \begin{equation}
            Q_{ij}^{(kq)} = 
            \begin{bmatrix}
                \bar{\vec{x}}_{\eta(q,i)}^T \left(\bar{A}_{kq}^T S_q + S_q \bar{A}_{kq}\right) \bar{\vec{x}}_{\eta(q,j)} & \ast \\ 1 - \bar{\vec{x}}_{\eta(q,i)}^T S_q \bar{\vec{x}}_{\eta(q,j)} & -1
            \end{bmatrix},
        \end{equation}
        $\bar{A}_{kq}$ as defined in \eqref{eq:x_Akq}, $\bar{\Lambda}_i = \Lambda_i \otimes I_2$, and
        \begin{align}
            &\bar{Z}_q^T\left(\bar{A}_{k\nu(q)_s}^T S_{\nu(q)_{\bar{s}}} + S_{\nu(q)_{\bar{s}}} \bar{A}_{k\nu(q)_s}\right)\bar{Z}_q \\&+ H_{skq} T_i + T_i^T H_{skq}^T + E^2_{skq} < 0 \\
        \end{align}
        for every $\vec{\delta}_q \in \mathcal{X}_g$, and
        \begin{align}
            \dfrac{1}{2}\left(\bar{Y}_{s}^{(kq)} + \left.\bar{Y}_{s}^{(kq)}\right.^T \right) + \bar{H}_{skq} \bar{T}_i + \bar{T}_i^T \bar{H}_{skq}^T + \bar{E}_{skq} \leq 0
        \end{align}
        for every $ \vec{\delta}_q \in \mathcal{X}\setminus\partial\mathcal{X}$, and
        \begin{align}
            \bar{Z}_q^T\left( S_{\nu(q)_1} - \begin{bmatrix} \vec{0} & \vec{0} \\ \vec{0} & 1\end{bmatrix}\right)\bar{Z}_q + G_{q} T_i + T_i^T G_{q}^T - E^1_{q} \geq 0
        \end{align}
        for every $\vec{\delta}_q \in \partial \mathcal{X}$, with $i \in \mathcal{I}_{n}$, $k \in \mathcal{I}_{r}$, $s \in \{1,2\}$, 
        \begin{equation}
            \bar{s} = \left\{ \begin{array}{ll} 2, &\textrm{if } s=1 \\ 1, &\textrm{if } s=2\end{array}\right.,
        \end{equation}
        \begin{equation}
            \bar{Z}_q = \begin{bmatrix} \vec{x}_{\eta_f(q,1)} & \dots & \vec{x}_{\eta_f(q,n)} \\ 1 & \dots & 1 \end{bmatrix},
        \end{equation}
        \begin{equation}
            \bar{Y}_s^{(kq)} = \begin{bmatrix}
                Y_{11s}^{(kq)} & \dots & Y_{1(n+1)s}^{(kq)} \\
                \vdots & \ddots & \vdots \\
                Y_{(n+1)1s}^{(kq)} & \dots & Y_{(n+1)(n+1)s}^{(kq)}
            \end{bmatrix}, 
        \end{equation}
        \begin{align}
            &Y_{ijs}^{(kq)} =\begin{bmatrix}
                \bar{\vec{x}}_{\eta_f(q,i)}^T \left(\bar{A}_{k\nu(q)_s}^T S_{\nu(q)_{\bar{s}}} + S_{\nu(q)_{\bar{s}}} \bar{A}_{k\nu(q)_s}\right) \bar{\vec{x}}_{\eta_f(q,j)} & \ast \\ 1 - \bar{\vec{x}}_{\eta_f(q,i)}^T S_{\nu(q)_{\bar{s}}} \bar{\vec{x}}_{\eta_f(q,j)} & -1
            \end{bmatrix},
        \end{align}
        \begin{equation}
            \bar{T}_i = T_i \otimes I_2,
        \end{equation}
        then the Lyapunov function in \eqref{eq:switch_V} is a guaranteed \emph{harmonically transformed} upper-bound time for the control law \eqref{eq:K}.
    \end{theorem}

    \begin{proof}
        Our goal is to minimize the integral $\int_\mathcal{X} \bar{V}(\vec{x})\vec{dx}$, in order to ensure that we are minimizing the upper bound in Theorem \ref{thm:harmonic_lyap} for all of the states. Note that we have
        \begin{equation}
            \int_\mathcal{X} \bar{V}(\vec{x})\vec{dx} = \sum_{q=1}^{r_s} \int_{\vec{\Delta}_q} V_q(\vec{x})\vec{dx} = \sum_{q=1}^{r_s} \int_{\vec{\Delta}_q} \bar{\vec{x}}^T S_q \bar{\vec{x}} \vec{dx},
        \end{equation}
        and from \citep[Theorem 2.1]{Lasserre2001}, we can write that
        \begin{align}
            &\sum_{q=1}^{r_s} \int_{\vec{\Delta}_q} \bar{\vec{x}}^T S_q \bar{\vec{x}} \vec{dx} = \sum_{q=1}^{r_s} \dfrac{\operatorname{vol}(\vec{\Delta}_q)}{\begin{pmatrix}n+2 \\ 2\end{pmatrix}}\left(\sum_{i=1}^{n+1} \sum_{j=i}^{n+1} \bar{\vec{x}}_{\eta(q,i)}^T S_q \bar{\vec{x}}_{\eta(q,j)}\right) \\&\;= \dfrac{2}{(n+1)(n+2)n!}\sum_{q=1}^{r_s} \left|\det\left(\bar{X}_q\right)\right|\left(\sum_{i=1}^{n+1} \sum_{j=i}^{n+1} \bar{\vec{x}}_{\eta(q,i)}^T S_q \bar{\vec{x}}_{\eta(q,j)}\right),
        \end{align}
        which matches the objective function presented in \eqref{eq:int_numerical}. From the conditions in Theorem \ref{thm:harmonic_lyap}, we have that the conditions for $\vec{x} \in \mathcal{X}\setminus\mathcal{X}_g$ can be written, using Schur's complement as
        \begin{align}
            \begin{bmatrix}
                \dot{\bar{V}}(\vec{x}) & 1 - \bar{V}(\vec{x}) \\ \ast & -1 
            \end{bmatrix} &\leq 0,
        \end{align}
        which, if we consider the closed loop system's dynamics and the representation chosen for the Lyapunov function, leads to
        \begin{align}
            \begin{bmatrix}
                \bar{\vec{x}}^T\left(\bar{A}_{\alpha q}^T S_q + S_q \bar{A}_{\alpha q}\right)\bar{\vec{x}}  & \ast \\ 1 - \bar{\vec{x}}^T S_q \bar{\vec{x}} & -1 
            \end{bmatrix} &\leq 0, \textrm{ for } \vec{x} \in \vec{\Delta}_q, \\
            \sum_{i=1}^{n+1} \sum_{j=1}^{n+1} \sum_{k = 1}^r \beta_i \beta_j \alpha_k Q_{ij}^{(kq)} &\leq 0, \textrm{ for }  \vec{\Delta}_q \in \mathcal{X}\setminus\mathcal{X}_g. \label{eq:sum_q}
        \end{align}

        In order to deal with this double summation in $\vec{\beta}$, we will make use of the structural relaxation approach \citep{kim2024,Campos2025b}. Note that this double summation can be rewritten as
        \begin{align}
            \sum_{k = 1}^r \alpha_k \left(\vec{\beta}^T \otimes I_2\right) \bar{Q}^{(kq)} \left(\vec{\beta} \otimes I_2\right) &\leq 0, \textrm{ for } \vec{\Delta}_q \in \mathcal{X}\setminus\mathcal{X}_g,\\
        \end{align}
        and since $\Lambda_\beta \vec{\beta} = 0 \Rightarrow (\Lambda_\beta \otimes I_2)(\vec{\beta} \otimes I_2) = 0$, then $\bar{N}_{\alpha q} \bar{\Lambda}_\beta (\vec{\beta} \otimes I_2) = 0$, and, given a nonnegative matrix $\bar{W}_{\alpha q}$, we have that $(\vec{\beta}^T \otimes I_2)\bar{W}_{\alpha q}(\vec{\beta} \otimes I_2) \geq 0$, we get that a sufficient condition is given by
        \begin{align}
            &\sum_{k = 1}^r \alpha_k \left(\vec{\beta}^T \otimes I_2\right) \Xi_{kq\beta} \left(\vec{\beta} \otimes I_2\right) \leq 0, \textrm{ for } \vec{\Delta}_q \in \mathcal{X}\setminus\mathcal{X}_g,
        \end{align}
        with 
        \begin{align}
            \Xi_{kq\beta} = \dfrac{1}{2}\left(\bar{Q}^{(kq)} + \left.\bar{Q}^{(kq)}\right.^T \right) + \bar{N}_{k q} \bar{\Lambda}_\beta + \bar{\Lambda}_\beta^T\bar{N}_{k q}^T + \bar{W}_{k q} . \end{align}

        Note that the condition $\dot{\bar{V}}(\vec{x}) <0$ for $\vec{x} \in \mathcal{X}_g$, can be written as
        \begin{align}
            \bar{\vec{x}}^T\left(\bar{A}_{\alpha q}^T S_q + S_q \bar{A}_{\alpha q}\right)\bar{\vec{x}} < 0, \textrm{ for } \vec{x} \in \vec{\Delta}_q \in \mathcal{X}_g,
        \end{align}
        and even though we could employ the S-procedure \citep{Johansson1998} as usual in this condition, we once again make use of the structural relaxation approach \citep{kim2024,Campos2025b}, by rewriting it as
        \begin{equation}
            \vec{\beta}^T \bar{X}_q^T \left(\bar{A}_{\alpha q}^T S_q + S_q \bar{A}_{\alpha q}\right) \bar{X}_q \vec{\beta} \leq 0, \textrm{ for } \vec{\Delta}_q \in \mathcal{X}_g,
        \end{equation}
        which by using a similar argument as the one presented above, leads to the sufficient condition
        \begin{equation}
            \vec{\beta}^T \Upsilon_{q\alpha} \vec{\beta} < 0, \;\forall \Delta_q \in \mathcal{X}_g,
        \end{equation}
        with
        \begin{align}
            \Upsilon_{q\alpha} = \bar{X}_q^T\left(\bar{A}_{\alpha q}^T S_q + S_q \bar{A}_{\alpha q}\right)\bar{X}_q + N_{\alpha q} \Lambda_\beta + \Lambda_\beta^T N_{\alpha q}^T + W^ 2_{\alpha q} . \end{align}

        In addition to this, in order to ensure that $\bar{V}(\vec{x})$ is positive definite, we impose that
        \begin{align}
            &\bar{\vec{x}}^T\left(S_q - \gamma \begin{bmatrix} I_n & \vec{0} \\\vec{0} & \vec{0} \end{bmatrix} \right)\bar{\vec{x}} \geq 0, \textrm{ for } \vec{x} \in \vec{\Delta}_q \in \mathcal{X}_g, \\
            &\bar{\vec{x}}^T S_q \bar{\vec{x}} \geq 0, \textrm{ for } \vec{x} \in \vec{\Delta}_q \in \mathcal{X}\setminus\mathcal{X}_g,
        \end{align}
        which, by making use of the structural relaxation approach, similarly leads to the sufficient conditions
        \begin{align}
            &\bar{X}_q^T\left(S_q - \gamma \begin{bmatrix} I_n & \vec{0} \\\vec{0} & \vec{0} \end{bmatrix} \right)\bar{X}_q + M_{q} \Lambda_{\beta} + \Lambda_{\beta}^T M_{q}^T - W^ 1_{q} \geq 0, 
        \end{align}    
        for every $\Delta_q \in \mathcal{X}_g$, and
        \begin{align}
            &\bar{X}_q^T S_q \bar{X}_q + M_{q} \Lambda_{\beta} + \Lambda_{\beta}^T M_{q}^T - W^ 1_{q} \geq 0, \;\forall \Delta_q \in \mathcal{X}\setminus\mathcal{X}_g.\\
        \end{align}
        
        In order to impose that $\bar{V}(\vec{x})\geq 1, \forall \vec{x} \in \partial \mathcal{X}$, we can make use of the structural relaxation approach once more, but this time considering the simplicial faces $\vec{\delta}_q$ which are on the boundary of the analysis region. Similarly to the approach employed before, note that this condition can be written as
        \begin{equation}
            \bar{\vec{\beta}}^T\bar{Z}_q \left( S_{\nu(q)_1} - \begin{bmatrix} \vec{0} & \vec{0} \\ \vec{0} & 1\end{bmatrix}\right)\bar{Z}_q \bar{\vec{\beta}} \geq 0, \;\forall \vec{\delta}_q \in \partial \mathcal{X},
        \end{equation}
        and using the fact that $T_{\bar{\beta}} \bar{\vec{\beta}} = 0$ and $\bar{\vec{\beta}}^T E^1_q \bar{\vec{\beta}} \geq 0$, for a nonnegative matrix $E^1_{q}$, we can obtain the sufficient conditions
        \begin{align}
            \bar{Z}_q^T\left( S_{\nu(q)_1} - \begin{bmatrix} \vec{0} & \vec{0} \\ \vec{0} & 1\end{bmatrix}\right)\bar{Z}_q + G_{q} T_{\bar{\beta}} + T_{\bar{\beta}}^T G_{q}^T - E^1_{q} \geq 0, \;\forall \vec{\delta}_q \in \partial \mathcal{X}.
        \end{align}
        Finally, since we do not impose continuity of the control law over the boundaries between simplices (given by the intersecting faces), we may have possible discontinuities on the closed loop dynamics over these boundaries, and need to take them into consideration. To that regard, we need to consider that over these boundaries the system's dynamics can be described by the convex hull of the dynamics over the two intersecting regions. In order to do so, the conditions over the boundaries need to be verified by employing the closed loop dynamics of one region with the Lyapunov function of the other region, leading to the conditions
        \begin{align}
            &\bar{\vec{x}}^T \left(\bar{A}_{\alpha\nu(q)_2}^T S_{\nu(q)_1} + S_{\nu(q)_1} \bar{A}_{\alpha\nu(q)_2}\right) \bar{\vec{x}} < 0, \\
            &\bar{\vec{x}}^T \left(\bar{A}_{\alpha\nu(q)_1}^T S_{\nu(q)_2} + S_{\nu(q)_2} \bar{A}_{\alpha\nu(q)_1}\right) \bar{\vec{x}} < 0,
        \end{align}
        for cases in which $\vec{x} \in \vec{\delta}_q \in \mathcal{X}_g$, and
        \begin{align}
            &\begin{bmatrix}
                \bar{\vec{x}}^T\left(\bar{A}_{\alpha \nu(q)_2}^T S_{\nu(q)_1} + S_{\nu(q)_1} \bar{A}_{\alpha \nu(q)_2}\right)\bar{\vec{x}}  & \ast \\ 1 - \bar{\vec{x}}^T S_{\nu(q)_1} \bar{\vec{x}} & -1 
            \end{bmatrix} \leq 0, \\
            &\begin{bmatrix}
                \bar{\vec{x}}^T\left(\bar{A}_{\alpha \nu(q)_1}^T S_{\nu(q)_2} + S_{\nu(q)_2} \bar{A}_{\alpha \nu(q)_1}\right)\bar{\vec{x}}  & \ast \\ 1 - \bar{\vec{x}}^T S_{\nu(q)_2} \bar{\vec{x}} & -1 
            \end{bmatrix} \leq 0,
        \end{align}
        for the cases in which $\vec{x} \in \vec{\delta}_q \in \mathcal{X} \setminus \mathcal{X}_g$. These conditions, in turn, lead to the remaining sufficient conditions presented in the theorem by making use of the structural relaxation approach one last time.
    \end{proof}

    \begin{theorem}[Policy Improvement] \label{thm:policy_improvement}
    Consider the uncertain polytopic system given by \eqref{eq:sys}, and a guaranteed \emph{harmonically transformed} upper-bound time function given by \eqref{eq:switch_V}. Consider that there are also matrix functions $\Lambda_\beta \in \mathbb{R}^{\frac{n(n+1)}{2} \times (n+1)}$ and $T_{\bar{\beta}} \in \mathbb{R}^{\frac{n(n-1)}{2} \times n}$ such that $\Lambda_\beta \vec{\beta} = 0, T_{\bar{\beta}} \bar{\vec{\beta}} = 0$. If there exist vectors $\vec{f}_\ell \in \mathbb{R}^{n+1}$, non-negative symmetric matrices $W^1_{q}, W^2_{kq} \in \mathbb{R}^{(n+1) \times (n+1)}, \bar{W}_{kq} \in \mathbb{R}^{2(n+1) \times 2(n+1)}, E^1_q, E^2_{skq} \in \mathbb{R}^{n \times n}, \bar{E}_{skq} \in \mathbb{R}^{2n \times 2n}$, matrices $N_{kq}, M_{q} \in \mathbb{R}^{(n+1) \times \frac{n(n+1)}{2}}, \allowbreak \bar{N}_{kq} \in \mathbb{R}^{2(n+1) \times n(n+1)}, G_q, H_{skq} \in \mathbb{R}^{n \times \frac{n(n-1)}{2}}, \bar{H}_{skq} \in \mathbb{R}^{2n \times n(n-1)}$, and matrices $\bar{K}_q \in \mathbb{R}^{m \times (n+1)}$, with $\vec{k}_{q} = 0$ if $\vec{\Delta}_q \in \mathcal{X}_g$, that maximize
    {\begin{equation}
        \dfrac{2}{(n+1)(n+2)n!}\left(\sum_{q = 1}^{r_s}\left|\det(\bar{X}_q)\right|\left(\sum_{i=1}^{n+1} \sum_{j=i}^{n+1} \bar{\vec{x}}_{\eta(q,i)}^T R_q \bar{\vec{x}}_{\eta(q,j)}\right)\right)
    \end{equation}}
    subject to
    \begin{align}
      &\bar{X}_q^T R_q \bar{X}_q + M_{q} \Lambda_i + \Lambda_i^T M_{q}^T - W^ 1_{q} \geq 0, \\
      &\bar{K}_q \bar{X}_q - \left(\begin{bmatrix}\underline{u}_1 & \dots & \underline{u}_m \end{bmatrix}^T\otimes\vec{1}^T_{n+1}\right) \succeq 0, \label{eq:const_u_low} \\
      &\bar{K}_q \bar{X}_q - \left(\begin{bmatrix}\bar{u}_1 & \dots & \bar{u}_m \end{bmatrix}^T\otimes\vec{1}^T_{n+1}\right) \preceq 0 \label{eq:const_u_high}
    \end{align}
    for every $\Delta_q \in \mathcal{X}$,
    \begin{align}
        \bar{X}_q^T\left(\bar{A}_{kq}^T S_q + S_q \bar{A}_{kq} + R_q\right)\bar{X}_q + N_{kq} \Lambda_i + \Lambda_i^T N_{kq}^T + W^ 2_{kq} < 0
    \end{align}
    for every $\Delta_q \in \mathcal{X}_g$,
    \begin{align}
        \dfrac{1}{2}\left(\bar{Q}^{(kq)} + \left.\bar{Q}^{(kq)}\right.^T \right) + \bar{N}_{kq} \bar{\Lambda}_i + \bar{\Lambda}_i^T \bar{N}_{kq}^T + \bar{W}_{kq} \leq 0
    \end{align}
    for every $\Delta_q \in \mathcal{X}\setminus\mathcal{X}_g$, with $i \in \mathcal{I}_{n+1}$, $k \in \mathcal{I}_{r}$,
    \begin{equation}
        R_q = \dfrac{1}{2}\left(F_q^T \bar{X}_q^{-1} + \bar{X}_q^{-T} F_q\right),
    \end{equation}
    in which $F_q \in \mathbb{R}^{(n+1)\times (n+1)}$ is a matrix whose columns corresponds to the $\vec{f}_\ell$ vectors related to each vertex composing simplex $\vec{\Delta}_q$, and
    \begin{equation}
        \bar{X}_q = \begin{bmatrix} \vec{x}_{\eta(q,1)} & \dots & \vec{x}_{\eta(q,n+1)} \\ 1 & \dots & 1 \end{bmatrix},
    \end{equation}
    \begin{equation}
        \bar{Q}^{(kq)} = \begin{bmatrix}
            Q_{11}^{(kq)} & \dots & Q_{1(n+1)}^{(kq)} \\
            \vdots & \ddots & \vdots \\
            Q_{(n+1)1}^{(kq)} & \dots & Q_{(n+1)(n+1)}^{(kq)}
        \end{bmatrix}, 
    \end{equation}
    \begin{equation}
        Q_{ij}^{(kq)} = 
        \begin{bmatrix}
            \bar{\vec{x}}_{\eta(q,i)}^T \left(\bar{A}_{kq}^T S_q + S_q \bar{A}_{kq} + R_q\right) \bar{\vec{x}}_{\eta(q,j)} & \ast \\ 1 - \bar{\vec{x}}_{\eta(q,i)}^T S_q \bar{\vec{x}}_{\eta(q,j)} & -1
        \end{bmatrix},
    \end{equation}
    $\bar{A}_{kq}$ as defined in \eqref{eq:x_Akq}, $\bar{\Lambda}_i = \Lambda_i \otimes I_2$ and
    \begin{align}
        &\bar{Z}_q^T\left(\bar{A}_{k\nu(q)_s}^T S_{\nu(q)_{\bar{s}}} + S_{\nu(q)_{\bar{s}}} \bar{A}_{k\nu(q)_s} + R_q\right)\bar{Z}_q \\&+ H_{1kq} T_i + T_i^T H_{1kq}^T + E^2_{skq} < 0
    \end{align}
    for every $\vec{\delta}_q \in \mathcal{X}_g$, and
    \begin{align}
        \dfrac{1}{2}\left(\bar{Y}_{s}^{(kq)} + \left.\bar{Y}_{s}^{(kq)}\right.^T \right) + \bar{H}_{skq} \bar{T}_i + \bar{T}_i^T \bar{H}_{skq}^T + \bar{E}_{skq} \leq 0
    \end{align}
    for every $ \vec{\delta}_q \in \mathcal{X}\setminus\partial\mathcal{X}$, with $i \in \mathcal{I}_{n}$, $k \in \mathcal{I}_{r}$, $s \in \{1,2\}$, 
    \begin{equation}
        \bar{s} = \left\{ \begin{array}{ll} 2, &\textrm{if } s=1 \\ 1, &\textrm{if } s=2\end{array}\right.,
    \end{equation}
    \begin{equation}
        \bar{Z}_q = \begin{bmatrix} \vec{x}_{\eta_f(q,1)} & \dots & \vec{x}_{\eta_f(q,n)} \\ 1 & \dots & 1 \end{bmatrix},
    \end{equation}
    \begin{equation}
        \bar{Y}_s^{(kq)} = \begin{bmatrix}
            Y_{11s}^{(kq)} & \dots & Y_{1(n+1)s}^{(kq)} \\
            \vdots & \ddots & \vdots \\
            Y_{(n+1)1s}^{(kq)} & \dots & Y_{(n+1)(n+1)s}^{(kq)}
        \end{bmatrix}, 
    \end{equation}
    {
    \begin{align}
        &Y_{ijs}^{(kq)} = \begin{bmatrix}
            \bar{\vec{x}}_{\eta_f(q,i)}^T \left(\bar{A}_{k\nu(q)_s}^T S_{\nu(q)_{\bar{s}}} + S_{\nu(q)_{\bar{s}}} \bar{A}_{k\nu(q)_s} + R_q\right) \bar{\vec{x}}_{\eta_f(q,j)} & \ast \\ 1 - \bar{\vec{x}}_{\eta_f(q,i)}^T S_{\nu(q)_{\bar{s}}} \bar{\vec{x}}_{\eta_f(q,j)} & -1
        \end{bmatrix},
    \end{align}}
    \begin{equation}
        \bar{T}_i = T_i \otimes I_2,
    \end{equation}
    then the control law given by \eqref{eq:K} maximizes the decay of the guaranteed \emph{harmonically transformed} upper-bound time given by \eqref{eq:switch_V} over the state, while guaranteeing that $\vec{u} \in \mathcal{U}$.
    \end{theorem}

    \begin{proof}
        Consider a piecewise quadratic function given by
        \begin{equation}
            \mathcal{F}(\vec{x}) = \bar{\vec{x}}^T R_q \bar{\vec{x}}, \textrm{ for } \vec{x} \in \vec{\Delta}_q
        \end{equation}
        defined, similarly to the Lyapunov function, over the triangulation of the state space. Following similar arguments (employing the structural relaxation approach) as in the proof of Theorem \ref{thm:policy_evaluation}, it is straightforward to show that the conditions of Theorem \ref{thm:policy_improvement} are sufficient for
        \begin{equation}
            \mathcal{F}(\vec{x}) \geq 0,
        \end{equation}
        and
        \begin{align}
            &\dot{\bar{V}}(\vec{x}) \leq -\mathcal{F}(\vec{x}), \textrm{ for } \vec{x} \in \mathcal{X}_g, \\
            &\dot{\bar{V}}(\vec{x})  + \left(1 - \bar{V}(\vec{x})\right)^2\leq -\mathcal{F}(\vec{x}), \textrm{ for } \vec{x} \in \mathcal{X}\setminus\mathcal{X}_g, \\
        \end{align}
        both over the simplices $\vec{\Delta}_q$ and the neighboring faces $\vec{\delta}_q$. Since the objective cost can be seen as maximizing $\int_\mathcal{X} \mathcal{F}(\vec{x}) \vec{dx}$, the optimization problem can be seen as searching for a control law that minimizes $\dot{\bar{V}}(\vec{x})$ over $\vec{x} \in \mathcal{X}_g$, and $\dot{\bar{V}}(\vec{x})  + \left(1 - \bar{V}(\vec{x})\right)^2$ over $\vec{x} \in \mathcal{X}\setminus\mathcal{X}_g$, and, therefore, the control law given by \eqref{eq:K} maximizes the decay of the guaranteed \emph{harmonically transformed} upper-bound time given by \eqref{eq:switch_V} over the state space. Finally, equations \eqref{eq:const_u_low} and \eqref{eq:const_u_high} ensure that, if $\vec{x} \in \Delta_q$, then $\bar{K}_q \bar{\vec{x}} \in \mathcal{U}$.
    \end{proof}

    \begin{remark} \label{rem:switched}
        It is straightforward to adapt the conditions presented for the guaranteed-time control of uncertain piecewise polytopic systems, described by
        \begin{align}
            \dot{\vec{x}} = \sum_{i=1}^r \alpha_i^{(q)} \left(A_i^{(q)}\vec{x} + B_i^{(q)}\vec{u}\right), \textrm{ for } \vec{x} \in \vec{\Delta}_q
        \end{align}
        in which a different system can be considered for each simplex. The only difference in the conditions would be that, instead of $\bar{A}_{kq}$ as defined in \eqref{eq:x_Akq}, we would have
        \begin{align}
            \bar{A}_{kq} = \begin{bmatrix}
                A_k^{(q)} +  B_{k}^{(q)}K_q &  B_k^{(q)} \vec{k}_q \\ \vec{0} & \vec{0}
            \end{bmatrix}.
        \end{align}
        In the same manner, these conditions could also be simply modified to deal with nonlinear systems by using a Takagi-Sugeno (TS) fuzzy model \citep{Takagi1985} to exactly represent the nonlinear system inside of each simplex. By making use of the sector nonlinearity approach \citep{Tanaka2001}, any input-affine nonlinear system, whose dynamics are Lipschitz continuous, and whose origin is an equilibrium point, can be written in the piecewise-TS form
        \begin{align}
            &\dot{\vec{x}} = \sum_{i=1}^r h_i^{(q)}(\vec{x}) \left(A_i^{(q)}\vec{x} + B_i^{(q)}\vec{u}\right), \textrm{ for } \vec{x} \in \vec{\Delta}_q, \\
            &\sum_{i=1}^r h_i^{(q)}(\vec{x}) = 1, \quad h_i^{(q)}(\vec{x}) \geq 0 \;\forall i.
        \end{align}
        Since the $h_i^{(q)}(\vec{x})$ membership functions are themselves convex weights, like the polytpic weights $\alpha_i^{(q)}$, the same conditions used for the uncertain piecewise polytopic systems can be employed in this case. Note that, unlike the $\alpha_i^{(q)}$ convex weights, the $h_i^{(q)}(\vec{x})$ membership functions are exactly known and could be employed in the control law and Lyapunov functions, but this approach \emph{will not be pursued in this paper}.
    \end{remark}

    \begin{remark}
        The conditions in Theorems \ref{thm:policy_evaluation} and \ref{thm:policy_improvement} require the use of matrix functions $\Lambda_\beta$ and $T_{\bar{\beta}}$. Though many different strategies could be employed to find these functions (such as the strategy proposed in \cite{kim2024}), in this paper we make use of the strategy proposed in \cite{Campos2025b}, described in the following (for the $\Lambda_\beta$ matrix) for convenience. Let $\hat{\vec{\beta}}^i$ be a reduced form of the vector $\vec{\beta}$, such that:
        \begin{equation}
            \hat{\vec{\beta}}_i = 
            \begin{bmatrix}
            \beta_{i}& \beta_{i+1} & \cdots & \beta_{n+1}    
            \end{bmatrix}^T
        \end{equation}
        
        Then the structural matrix can be constructed by the following rule:
        \begin{equation}\label{eq:rule_lambda}
            \Lambda_\beta = 
            \left[
            \begin{array}{c|c|c}
                \hat{\vec{\beta}}_2 & \multicolumn{2}{c}{-\beta_1 I_{n}} \\
                \hline
                0_{(n-1)\times 1} & \hat{\vec{\beta}}_3 & -\beta_2I_{n-1}\\
                \hline
                \vdots & \vdots & \vdots \\
                \hline
                0_{(n+1-i+1)\times (i-1)} & \hat{\vec{\beta}}_i & -\beta_{i-1}I_{n+1-i+1}\\
                \hline
                \vdots & \vdots & \vdots \\
                \hline
                0_{1\times (n-1)} & \hat{\vec{\beta}}_{n+1} & -\beta_{n}
            \end{array}
            \right]
        \end{equation}
        \noindent which satisfies the constraint $\Lambda_\beta \vec{\beta} = 0$.
    \end{remark}

\section{Examples}

In this section we present 3 examples to illustrate the proposed approach. Throughout them, the grid of points was chosen so that the region around the origin would be reasonably small and the proposed approach would steer the system towards the origin fast. In addition to this, $\gamma$ was chosen as a small value to ensure the positivity of the Lyapunov function inside of $\mathcal{X}_g$ while not affecting much the results outside of it. All examples were implemented on MATLAB R2025b using YALMIP as the parser and MOSEK as the optimization solver, using YALMIP optimizer interface to avoid the problem parsing overhead in the policy iteration, on a M4 Mac Studio with $32$ GB of RAM.

\subsection{Linear System - Double Tank System}
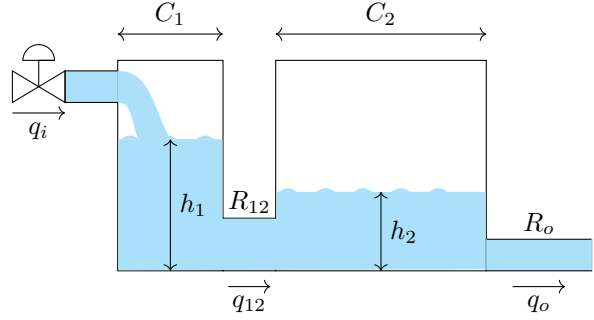
\begin{figure}
	\centering
	\begin{tikzpicture}[decoration={snake, segment length=5mm, amplitude=0.5mm},scale=0.7]
		\draw[black] (0,0) rectangle (2,4);
		\draw[black,<->] (0,4.5) -- (1,4.5) node[above]{$C_1$} --(2,4.5);
		\draw[black] (3,0) rectangle (7,4);
		\draw[black,<->] (3,4.5) -- (5,4.5) node[above]{$C_2$} -- (7,4.5);
		\draw[black] (2,0) rectangle (3,1);
		\node (R12) at (2.5,1.3) {$R_{12}$};
		\draw[black,->] (2.1,-0.3) -- (2.5,-0.3) node[below]{$q_{12}$} -- (2.9,-0.3);
		\fill[cyan!30] (0.01,0.01) -- (0.01,2.5) -- (1.99,2.5) -- (1.99, 0.99) -- (3.01, 0.99) -- (3.01, 1.5) -- (6.99,1.5) -- (6.99, 0.03) -- cycle;
		\fill[cyan!30,decorate]  (0.01,2.5) -- (1.99,2.5);
		\draw[black,<->] (1,0) -- (1,1.25) node[right]{$h_1$} -- (1,2.5);
		\fill[cyan!30,decorate]  (3.01, 1.5) -- (6.99,1.5);
		\draw[black,<->] (5,0) -- (5,0.75) node[right]{$h_2$} -- (5,1.5);
		\draw[black,fill=cyan!30] (-1,3.2) rectangle (0,3.8);
		\draw[cyan!30] (0.01,3.79) -- (0.01,3.21) -- (-0.01,3.21) -- (-0.01,3.79) -- cycle;
		\fill[cyan!30] (0,3.8) to[in=120,out=0](1,2.5) -- (0.4,2.5) to[in=0,out=120](0,3.2);
		\draw[black] (-1,3.2) -- (-2,3.8) -- (-2,3.2) -- (-1, 3.8) -- cycle;
		\draw[black] (-1.5,3.5) -- (-1.5, 4) -- (-1.2,4) -- (-1.8,4) arc[start angle=180, end angle=0, radius=0.3cm];
		\draw[black,->] (-2,3) -- (-1.5,3) node[below]{$q_i$} -- (-1,3);
		\draw[black,fill=cyan!30] (7,0) rectangle (9,0.6);
		\node (Ro) at (8,0.9) {$R_o$};
		\draw[black,->] (7.5,-0.3) -- (8,-0.3) node[below]{$q_o$} -- (8.5,-0.3);
		\draw[cyan!30] (7.01,0.01) -- (7.01,0.6) -- (6.99,0.6) -- (6.99,0.01) -- cycle;
		\draw[cyan!30] (9,0.01) -- (9,0.6);
	\end{tikzpicture}
	\caption{Double Tank System.}
	\label{fig:double_tank}
\end{figure}
Consider the double tank system illustrated in Figure \ref{fig:double_tank}. A simplified version of its dynamics can be described by
\begin{align}
	\dot{h}_1 &= \frac{-1}{R_{12}C_1}h_1 + \frac{1}{R_{12}C_1}h_2 + \frac{\overline{q}_i}{C_1}\delta, \\
	\dot{h}_2 &= \frac{1}{R_{12}C_2}h_1 - \left(\frac{R_{12} + R_o}{R_{12}R_o C_2}\right)h_2,
\end{align}
with $h_1$ and $h_2$ representing the level of both tanks, $C_1$ and $C_2$ are the area of the section of each tank, $\bar{q}_i$ represents the maximum input flow, $\delta \in [0,1]$ represents the opening of the input valve, and $R_{12}$ and $R_o$ are hydraulic resistances. Without loss of generality, we will consider that the problem is driving the level of the second tank to a desired equilibrium value $\bar{h}_2$, with a corresponding equilibrium level $\bar{h}_1$ for the first tank and equilibrium value for the input valve $\bar{\delta}$. With a small change of coordinates, such that this desired equilibrium point is the origin of the state space, the dynamics of the system are given by
\begin{align}
    \dot{\vec{x}} = \begin{bmatrix} \dfrac{-1}{R_{12}C_1} & \dfrac{1}{R_{12}C_1} \\ \dfrac{1}{R_{12}C_2} & - \left(\dfrac{R_{12} + R_o}{R_{12}R_o C_2}\right) \end{bmatrix} \vec{x} + \begin{bmatrix} \dfrac{\overline{q}_i}{C_1} \\ 0 \end{bmatrix} u,
\end{align}
with $\vec{x} = \begin{bmatrix} h_1 - \bar{h}_1 & h_2 - \bar{h}_2 \end{bmatrix}^T$ and $u = \delta - \bar{\delta}$. The parameters employed for this system were $C_1 = 2 m^2$, $C_2 = 4 m^2$, $\bar{q}_i = 0.4 m^3/s$, $R_{12} = 1 s/m^2$, $R_o = 10 s/m^2$, $\bar{h}_1 = 2.2 m$, $\bar{h}_2 = 2 m$, and $\bar{\delta} = 0.5$. In these new coordinates, we have that $x_1 \in [-2.2, 2.2], x_2 \in [-2, 2]$ and $u \in [-0.5, 0.5]$.

In order to solve this problem, we employ the Guaranteed Time Control approach proposed in Theorems \ref{thm:policy_evaluation} and \ref{thm:policy_improvement}, by making use of a regular grid of $15 \times 15$ points in the state space, leading to 225 points and 392 simplices. Since the system is open-loop stable, the initial controller (used to start the policy iteration) is chosen as not using any control ($K_q = 0 \; \forall q)$, and 7 iterations were performed (alternating between Theorems \ref{thm:policy_evaluation} and \ref{thm:policy_improvement}), considering $\gamma = 10^{-3}$.

\begin{figure}
	\centering
    \includegraphics{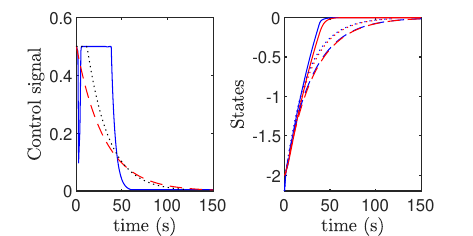}
    \caption{A comparison of the closed loop behavior for the double tank system, from the empty tanks initial condition, obtained from the Guaranteed Time Control  (solid lines), using Theorems \ref{thm:policy_evaluation} and \ref{thm:policy_improvement}, a linear control law that imposes the maximum decay rate without saturating the control input (dashed lines), and a saturated linear control law \citep[Adapted from Proposition 3.10]{Tarbouriech2011} that imposes the maximum decay rate while guaranteeing the largest estimated domain of attraction (dotted lines).}
    \label{fig:tank_compare_decay}
\end{figure}

\begin{figure}
	\centering
    \includegraphics{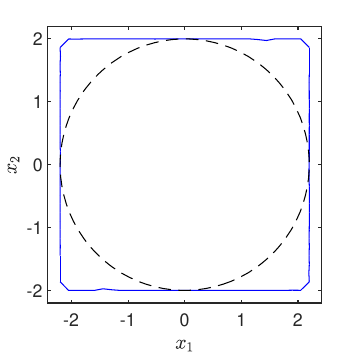}
    \caption{A comparison of the estimation of the closed loop's Domain of Attraction for the double tank system obtained from the Guaranteed Time Control approach (\textcolor{blue}{blue solid line}), using Theorems \ref{thm:policy_evaluation} and \ref{thm:policy_improvement} and a saturated linear control law \citep[Adapted from Proposition 3.10]{Tarbouriech2011} that imposes the maximum decay rate while guaranteeing the largest estimated domain of attraction (black dashed line).}
    \label{fig:tank_compare_DoA}
\end{figure}

The results were initially compared against a linear control law that imposes the maximum decay rate it can, without ever saturating inside of the analysis region and a saturated linear control law \citep[Adapted from Proposition 3.10]{Tarbouriech2011} that imposes the maximum decay rate while guaranteeing the largest estimated domain of attraction. This comparison is shown in Figure \ref{fig:tank_compare_decay}, for an initial condition that corresponds to both tanks completely empty, as well as in Figure \ref{fig:tank_compare_DoA}, which depicts the comparison of the estimates of the domains of attraction from the proposed conditions and the saturated linear control law \citep[Adapted from Proposition 3.10]{Tarbouriech2011}.

As shown in Figure \ref{fig:tank_compare_decay}, the proposed control law is capable of saturating the control signal for a significantly larger amount of time than the saturated linear control law, leading to a significantly faster time to reach the origin than both the linear and saturated linear control laws.  In addition to this, as shown in Figure \ref{fig:tank_compare_DoA}, the estimate of the domain of attraction of the origin in closed loop is considerably larger in our approach (consisting of almost all of the state space) compared against the estimate from the conditions in \citep[Proposition 3.10]{Tarbouriech2011} (which is the largest area a quadratic Lyapunov function could estimate in this example).

\begin{figure}
	\centering
    \includegraphics{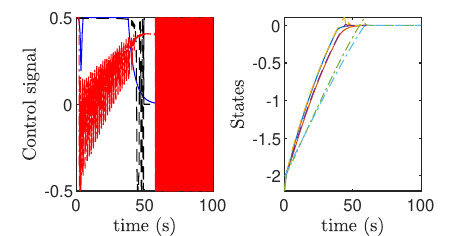}
    \caption{A comparison of the closed loop behavior for the double tank system, using the Guaranteed Time Control  approach (solid lines, \textcolor{blue}{blue control signal}) against a finite-time control law based on an Implicit Lyapunov function approach \citep{Polyakov2015} (dashed lines, black control signal), and a finite-time control law based on Prescribed Performance Control \citep{Guo2022} (dash-dotted lines, \textcolor{red}{red control signal}).}
    \label{fig:tank_compare_finte_time}
\end{figure}

In a different set of comparisons, regarding the behavior of the closed loop system starting from the empty tanks initial condition, the proposed approach was also compared against two different finite-time control approaches, one based on an implicit Lyapunov function \citep{Polyakov2015}, and another based on the idea of Prescribed Performance Control \citep{Guo2022}. The results are shown in Figure \ref{fig:tank_compare_finte_time}, with the performance functions used for the prescribed performance control tuned slightly slower in order for us to be able to set the responses apart.

As shown in the figure, the control law based on the proposed approach is capable of driving the system towards the desired goal with a similar closed loop behavior as the finite-time control approaches, however with a considerably smoother control action.

\begin{figure}
	\centering
    \includegraphics{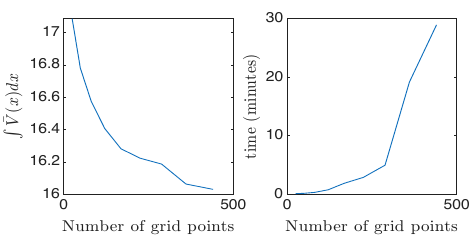}
    \caption{The effect of the grid size on the Guaranteed Time Control approach for the double tank system. The left plot shows the $\int_\mathcal{X} \bar{V}(x) dx$ (objective function) over the number of grid points, while the second plot show the time taken to find a solution over the number of grid points.}
    \label{fig:tank_compare_grid_time_cost}
\end{figure}

\begin{figure}
	\centering
    \includegraphics{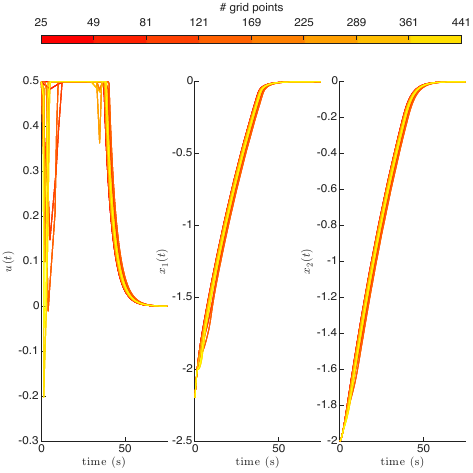}
    \caption{A comparison of the closed loop behavior for the double tank system, using the Guaranteed Time Control over different grid sizes. The first plot shows the control action, whereas the second and third plots show the evolution of the states over time.}
    \label{fig:tank_compare_grid_closed_loop}
\end{figure}

Finally, we also study the effect of varying the grid size in this problem. Instead of using a strictly uniform grid over the entire state space, the discretization was explicitly anchored around a predetermined inner boundary defined by $\varepsilon_{\textrm{goal}} = 0.3$. For a given odd number of points per axis $n_x$, the intervals $[-\bar{x}_i, -\varepsilon_{\textrm{goal}}]$ and $[\varepsilon_{\textrm{goal}}, \bar{x}_i]$, with $\bar{x}_1 = 2.2, \bar{x}_2 = 2$, were uniformly divided into $(n_x-1)/2$ points. The origin was inserted as a central point, and the final grid taken as the cartesian product of both one-dimensional grids. This led to a final grid with $n_x^2$ points and the same $\mathcal{X}_g$ region around the origin for all of the grids, allowing for comparisons regarding the upper bound to $\int_\mathcal{X} \bar{V}(x) dx$ (our objective function) and the time taken for the solution. In this test, we varied $n_x$ from $5$ to $21$, leading to grid sizes varying from $25$ to $441$, and ran the Policy Iteration until the cost variation was less than or equal to $10^{-2}$. Figure \ref{fig:tank_compare_grid_time_cost} shows the evolution of the objective function and the time taken to solve the problem with respect to the number of grid points, whereas Figure \ref{fig:tank_compare_grid_closed_loop} presents the closed loop behavior for all of the grid sizes plotted together. It can be clearly seen that while the objective function decreases monotonically with the number of grid points considered, the time taken to solve the problem increases monotonically, and defining a suitable grid (possibly non-regular) for the triangulation is an interesting direction for future research. Note also that, even though the cost function decreases with every increasing grid size, in this particular example, all of the controllers behaved quite similarly. This highlights an interesting aspect of the guaranteed time approach: even though we only ensure an upper bound for the value function, we can guarantee that it will work in closed loop (if a solution was found) and a coarse grid can still lead to good results.

\subsection{Unstable Uncertain Polytopic System}
In this example, we consider a numerical problem, in which we have an open loop unstable uncertain polytopic system composed by 2 vertices, described by \eqref{eq:sys} with matrices
\begin{align}
    &A_1 = \begin{bmatrix}
        -0.2868  & -3.5353 \\
        -3.5353  & -8.7132
    \end{bmatrix}, && 
    A_2 = \begin{bmatrix}
        -0.2868  & 3.5353 \\
        3.5353  & -8.7132
    \end{bmatrix},
\end{align}
$B_1 = B_2 = \begin{bmatrix} 1 & -1 \end{bmatrix}^T$. We consider that $x_1 \in [-5, 5]$, $x_2 \in [-5, 5]$ and $u \in [-10, 10]$.

Similarly to the previous example, the Guaranteed Time Control approach proposed in Theorems \ref{thm:policy_evaluation} and \ref{thm:policy_improvement} is employed, making use of a regular grid of $15 \times 15$ points in the state space, leading to 225 points and 392 simplices. Since the system is open-loop unstable, we make use of an initial controller that imposes a minimum decay rate of 1 to a quadratic Lyapunov function, while minimizing the controller gain (but ignoring the control bounds). This controller can be found by solving the optimization problem
\begin{align}
    &\min_{X,Y,s} s, \\
    s.t.\\
    &X\geq I, \\
    &A_iX+XA_i^T + BY+Y^TB^T + 2X \leq 0, \\
    &\begin{bmatrix}
        -s & Y \\ \ast &-X
    \end{bmatrix} \leq 0,
\end{align}
which leads to the initial gain $K = YX^{-1} = \begin{bmatrix} -3.2668 & -1.0985\end{bmatrix}$. After this initialization, 10 iterations were performed (alternating between Theorems \ref{thm:policy_evaluation} and \ref{thm:policy_improvement}), considering $\gamma = 10^{-3}$.

The results were compared against an adaptation of the Semi-Lagragian approximation of time-optimal control problems in \citep{Campos2025} to a zero-sum dynamic game setting in which one agent, controlling $\vec{\alpha}$, tries to maximize the time taken to reach the origin, and the other, controlling $u$, tries to minimize the time.

\begin{figure}
	\centering
    \includegraphics{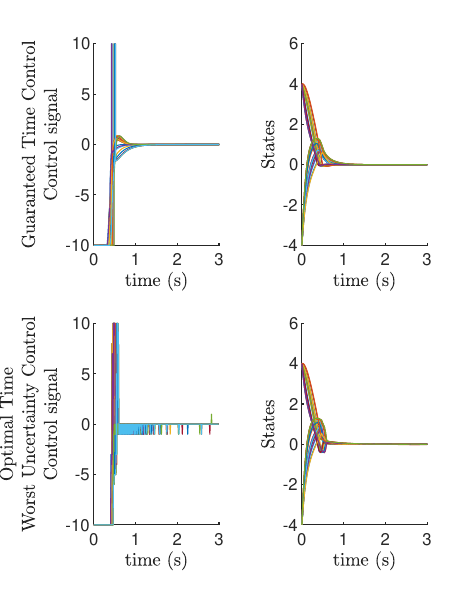}
    \caption[]{Closed loop behavior for the unstable uncertain polytopic system obtained from the Guaranteed Time Control approach, using Theorems \ref{thm:policy_evaluation} and \ref{thm:policy_improvement}, and from an optimal time worst case uncertainty control, starting from the initial condition $\begin{bmatrix}4 & -4\end{bmatrix}^T$, considering 20 different random instances for the values of $\vec{\alpha}$ (represented by the different colors in the plots).}
    \label{fig:uncertain_input_states}
\end{figure}

\begin{figure}
	\centering
    \includegraphics{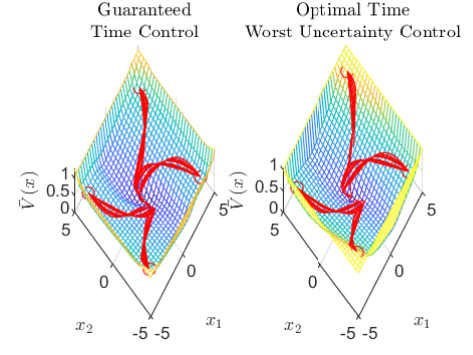}
    \caption[]{$\bar{V}(\vec{x})$ function for the unstable uncertain polytopic system obtained from the Guaranteed Time Control approach, using Theorems \ref{thm:policy_evaluation} and \ref{thm:policy_improvement}, and from an optimal time worst case uncertainty control. The red lines represent the closed loop trajectories from initial conditions $\begin{bmatrix}4 & -4\end{bmatrix}^T, \begin{bmatrix}-4 & 4\end{bmatrix}^T, \begin{bmatrix}-4 & -4\end{bmatrix}^T$ and $\begin{bmatrix}4 & 4\end{bmatrix}^T$, considering 20 different random instances for the values of $\vec{\alpha}$.}
    \label{fig:uncertain_value}
\end{figure}

This comparison is shown in Figures \ref{fig:uncertain_input_states} and \ref{fig:uncertain_value}, which show respectively the time evolution of the closed loop system, and the $\bar{V}(\vec{x})$ Lyapunov function as well as the behavior from 4 different initial conditions. As can be seen in the figures, the results obtained are comparable to the ones found from the optimal control approach (though slightly slower) with the added benefit that the control law found is already guaranteed to work in closed loop (since unlike the optimal control approach no approximation of the problem are done to solve it).

\subsection{Nonlinear System - Chua's circuit}
\begin{figure}
    \centering
    \begin{circuitikz}[american,scale=0.75]
        \draw (0,0) -- (0,0.75);
        \draw (-0.3,0.75) rectangle (0.3,2.25);
        \filldraw[fill=black] (-0.3,0.75) rectangle (0.3,1.0);
        \draw (0,0.75) to[R] (0,2.25);
        \draw (0,2.25) -- (0,3);
        \draw (0,3) node[pos=0.5,left,xshift=-1.5em,yshift=7.5em,rotate=90]{Chua's diode} -- (2,3) 
            to[C=$C_1$,v=$V_{C_1}$] (2,0) -- (0,0);
        \draw (2,0) -- (4,0) 
            to[cisource, l_=$u$] (4,3) -- (2,3);
        \draw (4,3) to [R=$R$] (7,3) to[C=$C_2$,v=$V_{C_2}$] (7,0) -- (2,0);
        \ctikzset{cute inductors}
        \draw (7,0) -- (9,0) to [L, l_=$L$,mirror,i=$i_L$] (9,3) -- (7,3);
    \end{circuitikz}
    \caption{Chua's circuit with one controlled current input}
    \label{fig:chua}
\end{figure}
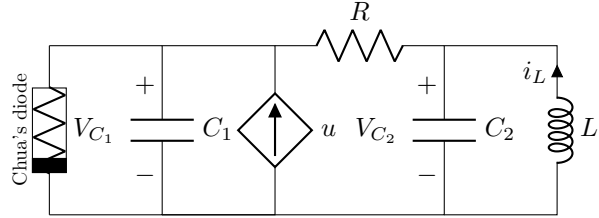

Consider the chaotic oscillator, known as Chua's circuit \citep{Chua1994}, with a controlled current source added illustrated in Figure \ref{fig:chua}. If we consider that $V_{C_1}$ and $V_{C_2}$ are expressed in $V$, whereas $i_L$ and $u$ are expressed in $mA$, it follows that the system can be described by
\begin{align}
    \dot{V}_{C_1} &= \frac{1}{C_1} \left(\frac{V_{C_2} - V_{C_1}}{R} - G(V_{C_1}) V_{C_1} + \frac{u}{1000}\right), \\
    \dot{V}_{C_1} &= \frac{1}{C_2} \left(\frac{V_{C_1} - V_{C_2}}{R} + \frac{i_L}{1000}\right), \\
    \frac{d}{dt} i_L &= \frac{-1000 V_{C_2}}{L}.
\end{align}
with $G(V_{C_1})$ the conductance of the Chua's diode, given by
\begin{align}
    G(V_{C_1}) = \left\{\begin{array}{ll} G_a, & \textrm{if } |V_{C_1}| < E \\ G_b + (G_a-G_b)\dfrac{E}{|V_{C_1}|}, & \textrm{otherwise}. \end{array}\right.
\end{align}
The parameters employed for this system were $C_1 = 30.14 \mu F$, $C_2 = 185.66 \mu F$, $L = 52.28 H$, $R = 1673 \Omega$, $G_a = -0.801 mS$, $G_b = -0.365 mS$, $E = 1.74 V$. We consider that $V_{C_1} \in [-6, 6]V$, $V_{C_2} \in [-3, 3]V$, $i_{L} \in [-3, 3]mA$ and $u \in [-200, 200]mA$.

Similarly to the previous examples, the Guaranteed Time Control approach proposed in Theorems \ref{thm:policy_evaluation} and \ref{thm:policy_improvement} is employed, making use of a regular grid of $11 \times 5 \times 5$ points in the state space, leading to 275 points and 960 simplices. Since we are dealing with a nonlinear system, for each simplex we employ a local Takagi-Sugeno model (as discussed in Remark \ref{rem:switched}) by considering the \emph{local} maximum and minimum values of $G(V_{C_1})$. Since the system is open-loop unstable, we make use of an initial linear stabilizing controller (that does not saturate in the region $\mathcal{X}_g$ around the origin) given by $K = \begin{bmatrix}-38.8017 & -31.1041 & 25.3298\end{bmatrix}$ and 15 iterations were performed (alternating between Theorems \ref{thm:policy_evaluation} and \ref{thm:policy_improvement}), considering $\gamma = 10^{-3}$.

\begin{figure}
	\centering
    \includegraphics{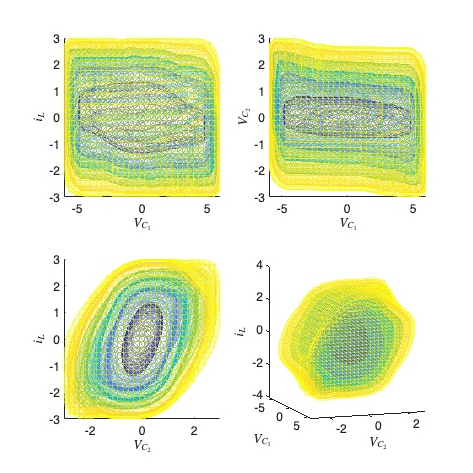}
    \caption{Level sets for the $\bar{V}(\vec{x})$ function for the Chua's circuit obtained from the Guaranteed Time Control approach, using Theorems \ref{thm:policy_evaluation} and \ref{thm:policy_improvement} represented as isosurfaces with values ranging from 0.1 to 0.99.}
    \label{fig:chua_iso}
\end{figure}

\begin{figure}
	\centering
    \includegraphics{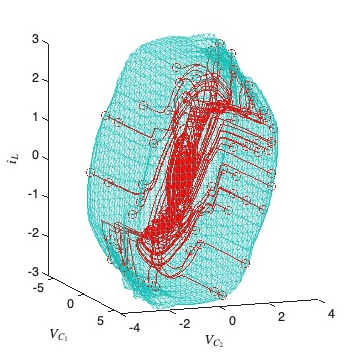}
    \caption{Estimation of the domain of attraction for the Chua's circuit, represeting the 0.99 level set of the $\bar{V}(\vec{x})$ function, obtained from the Guaranteed Time Control approach using Theorems \ref{thm:policy_evaluation} and \ref{thm:policy_improvement}. The red lines represent the closed loop trajectories from different initial conditions starting on the border of this level set.}
    \label{fig:chua_traj}
\end{figure}

\begin{figure}
	\centering
    \includegraphics{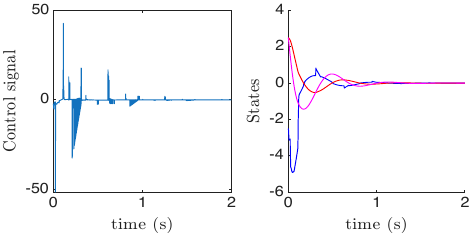}
    \caption[]{Closed loop behavior for the Chua's circuit from the Guaranteed Time Control approach, using Theorems \ref{thm:policy_evaluation} and \ref{thm:policy_improvement}, from the initial condition $\begin{bmatrix}-2.5 & 2.5 & 2.5 \end{bmatrix}^T$.}
    \label{fig:chua_cl}
\end{figure}

\begin{figure}
	\centering
    \includegraphics{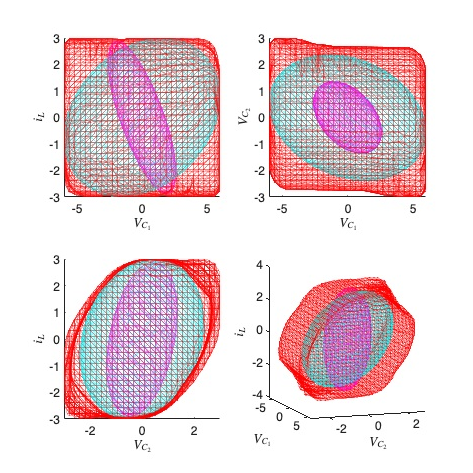}
    \caption{Comparison of the largest closed loop's domain of attraction estimation found for the Chua's circuit. The \textcolor{red}{red surface} represents the Guaranteed Time Control approach, using Theorems \ref{thm:policy_evaluation} and \ref{thm:policy_improvement}, whereas the \textcolor{cyan}{cyan surface} represents the largest closed loop's domain of attraction found using the approach in \cite[Theorem 2]{Lee2014}, and the \textcolor{magenta}{magenta surface} represents the largest closed loop's domain of attraction found using the approach in \cite[Algorithm 1]{Yara2025}.}
    \label{fig:chua_doa}
\end{figure}

The level sets for the Lyapunov function found for this system are illustrated in Figure \ref{fig:chua_iso}, whereas the 0.99 sublevel-set (an estimate of the closed loop's domain of attraction) is illustrated in Figure \ref{fig:chua_traj} along with several trajectories with initial conditions on the border of this set. In addition to this, Figure \ref{fig:chua_cl} presents the closed loop behavior of the system, starting from initial condition $\begin{bmatrix}-2.5 & 2.5 & 2.5 \end{bmatrix}^T$, and Figure \ref{fig:chua_doa} compares the estimate of the domain of attraction found using the Guaranteed Time Control approach against the estimate found from using the conditions in \cite[Theorem 2]{Lee2014} (modified to maximize the volume of an ellipsoidal volume inside of the domain of attraction, using $\max \operatorname{logdet}$ instead of maximizing the volume of a ball contained inside of the domain of attraction), with $|\dot{h}_i| \leq 40$, and the one found using the conditions in \cite[Algorithm 1]{Yara2025}.

Note that, similar to the other examples, the Guaranteed Time Control approach proposed in this paper was capable of finding a control law whose estimated domain of attraction covers most of the desired analysis region, and completely encompasses the estimate found from the competing approaches. Figure \ref{fig:chua_traj} shows that the estimate found is indeed a valid approximation of the domain of attraction as the system states converge towards the origin from the initial conditions picked on the border of the estimate found. Finally, Figure \ref{fig:chua_cl} show that the control bounds are respected, and that the extra conditions, imposed over the faces of the simplices, are indeed needed in this example since it exhibits a switching behavior in its control signal (likely due to discontinuities on the control law) but is still ensured to stabilize the system.

\section{Conclusion and Future Directions}

This paper proposed a Guaranteed Time Control approach for dynamical systems. In order to do so, Theorem \ref{thm:harmonic_lyap} proposed a set of Lyapunov conditions, which, if satisfied, ensures asymptotic stability, finds an upper-bound for the time taken to reach the desired set $\mathcal{X}_g$ that contains the origin, and ensures that the strict 1-sublevel set of the Lyapunov function is an estimate of the origin's domain of attraction. Theorems \ref{thm:policy_evaluation} and \ref{thm:policy_improvement} make use of these conditions, as well as the novel piecewise representation presented in Section \ref{sect:piece}, to propose a \emph{Policy Iteration} approach for the problem (circumventing the bilinear terms that appear if we try to solve for both Lyapunov function and control law simultaneously).

Three examples were presented to illustrate the proposed approach and compare it against the literature: a linear system, an unstable uncertain polytopic system, and a nonlinear system. The first example showed that the approach is capable of achieving considerably better results than simple linear conditions (both in terms of time taken to reach the origin, and on the guaranteed domain of attraction) and that it is capable of finding comparable results with finite time control laws (both based on ILF and PPC projects) while finding a considerably smoother control law. The second example illustrates that, given a dense enough discretization of the state space, the proposed approach is capable of finding results comparable to the solution of a time-optimal control law. The third example shows that the proposed approach is capable of handling highly nonlinear systems (given that it is a chaotic system in open-loop) and that the guaranteed domain of attraction is superior to the current state-of-the-art conditions.

Even though the proposed approach was capable of handling the examples presented in this paper, like many other approaches based on discretizing the state space in some manner, it suffers from the \emph{curse of dimensionality} with the number of simplices increasing exponentially as the dimension of the problem grows. In that regard, the adaptation of Domain Decomposition Methods \citep{Festa2018} and possible adaptive grids \citep{Munos2002} are very interesting research directions as they would allow for a decomposition/parallelization of the solution, as well as a partition with a reduced number of simplices. Other possible future directions include the use of control laws that incorporate further information about the system, either via \emph{Parallel Distributed Compensation} \citep{Tanaka2001} for the nonlinear systems, or adaptive polytopic controllers \citep{Campos2021} for the uncertain systems.

\bibliographystyle{apalike}
\bibliography{bibliography}
\end{document}